\documentclass[reprint,amsmath,amssymb,aps,prd]{revtex4-1}
\usepackage{graphicx}
\usepackage{dcolumn}
\usepackage{bm}
\usepackage{subfigure}

\begin{document}
\preprint{APS/123-QED}

\title{3D Lumped LC Resonators as Low Mass Axion Haloscopes}
\author{Ben T. McAllister}
\email{ben.mcallister@uwa.edu.au}
\author{Stephen R. Parker}
\affiliation{ARC Centre of Excellence for Engineered Quantum Systems, School of Physics, The University of Western Australia, 35 Stirling Highway, Crawley 6009, Western Australia, Australia}
\author{Michael E. Tobar}
\email{michael.tobar@uwa.edu.au}
\affiliation{ARC Centre of Excellence for Engineered Quantum Systems, School of Physics, The University of Western Australia, 35 Stirling Highway, Crawley 6009, Western Australia, Australia}
\date{\today}

\begin{abstract}
The axion is a hypothetical particle considered to be the most economical solution to the strong CP problem. It can also be formulated as a compelling component of dark matter. The haloscope, a leading axion detection scheme, relies on the conversion of galactic halo axions into real photons inside a resonant cavity structure in the presence of a static magnetic field, where the generated photon frequency corresponds to the mass of the axion. For maximum sensitivity it is key that the central frequency of the cavity mode structure coincides with the frequency of the generated photon. As the mass of the axion is unknown, it is necessary to perform searches over a wide range of frequencies. Currently there are substantial regions of the promising pre-inflationary low mass axion range without any viable proposals for experimental searches. We show that 3D resonant LC circuits with separated magnetic and electric fields, commonly known as re-entrant cavities, can be sensitive dark matter haloscopes in this region, with frequencies inherently lower than those achievable in the equivalent size of empty resonant cavity. We calculate the sensitivity and accessible axion mass range of these experiments, designing geometries to exploit and maximize the separated magnetic and electric coupling of the axion to the cavity mode.
\end{abstract}

\maketitle
\section{Introduction}
Of the numerous solutions to the strong CP problem in QCD, Peccei and Quinn's introduction of a new U(1) axial gauge symmetry is widely considered one of the most attractive. A consequence of the Peccei-Quinn solution is the generation of a pseudo-Goldstone boson, known as the axion~\cite{axion}. The axion is a type of weakly interacting slim particle, or WISP, which has been formulated as a component of cold dark matter~\cite{Sikivie1983,Preskill1983,Dine1983,cdm}. Consequently, the direct detection of axions is highly sought after, as their existence goes a long way to solving two of the largest problems in physics today. However, other than some broad limits from cosmological observation~\cite{limits}, many properties of the axion are currently unknown, presenting a serious barrier to detection.

The Sikivie haloscope~\cite{haloscope1,haloscope2} is one of the most sensitive and mature approaches to axion detection. The haloscope relies on the Primakoff effect, in which an axion is converted into a real photon, with a virtual photon present to conserve momentum. In a typical haloscope some kind of resonant cavity system is embedded in an external static magnetic field; if axions are present in the cavity some should spontaneously convert to photons, and this signal will be resonantly enhanced by the cavity structure if the photon frequency generated corresponds to the central frequency of a cavity mode. This signal can then be resolved above the thermal noise of the readout system. 

It has long been established that the power expected due to axion-photon conversion in a Sikivie haloscope is given by~\cite{haloscope1,haloscope2,ADMX2011}
\begin{equation}
\text{P}_{a}=\left(\frac{g_{\gamma}\alpha}{\pi f_{a}}\right)^{2}\frac{\rho_{a}}{m_{a}}VB_{0}^{2}\text{Q}\text{C}, \label{eq:paxion}
\end{equation}
Where $\text{g}_\gamma$ is an axion-model dependent parameter of $\mathcal{O}$(1)~\cite{KSVZ1,KSVZ2,DFSZ1},~$\text{f}_\text{a}$ is the Peccei-Quinn symmetry breaking scale, an important and unknown parameter in axion theory, $\rho_a$ the local density of axions, $m_a$ the mass of the axion - also unknown and dependent upon $\text{f}_\text{a}$, V is the volume of the resonant cavity, $\text{B}_0$ the strength of the external magnetic field, Q the loaded cavity quality factor (provided it is less than the quality factor of the axion signal, expected to be of order $10^6$) and C is a mode dependent form factor, which accounts for the degree of overlap between the cavity mode and the electric and magnetic field components of the photons generated by axion-photon conversion~\cite{OurPRL}. For maximum sensitivity in a haloscope it is important that the central frequency of the detecting cavity mode coincide with the frequency of the generated photons; this is assumed to be true in eq.~\eqref{eq:paxion}. The frequency of the generated photons depends upon the mass of the axion according to
\begin{equation}
E=hf_a=m_ac^2+\frac{1}{2}m_ac^2\beta^2,
\end{equation}
\begin{figure*}[t!]
	\subfigure[]{
		\includegraphics[width=0.49\textwidth]{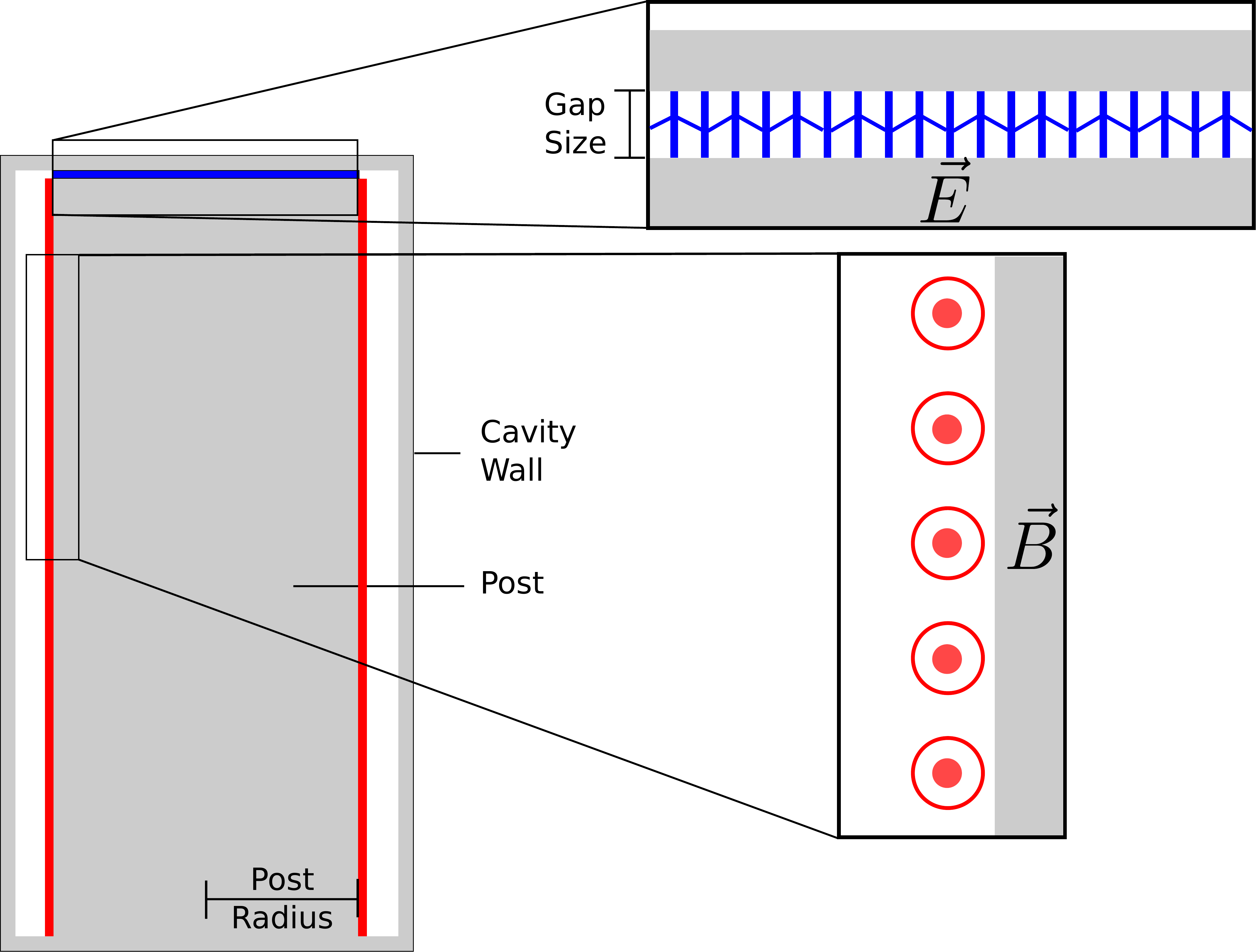}\label{fig:postcav}}
	\subfigure[]{
		\includegraphics[width=0.49\textwidth]{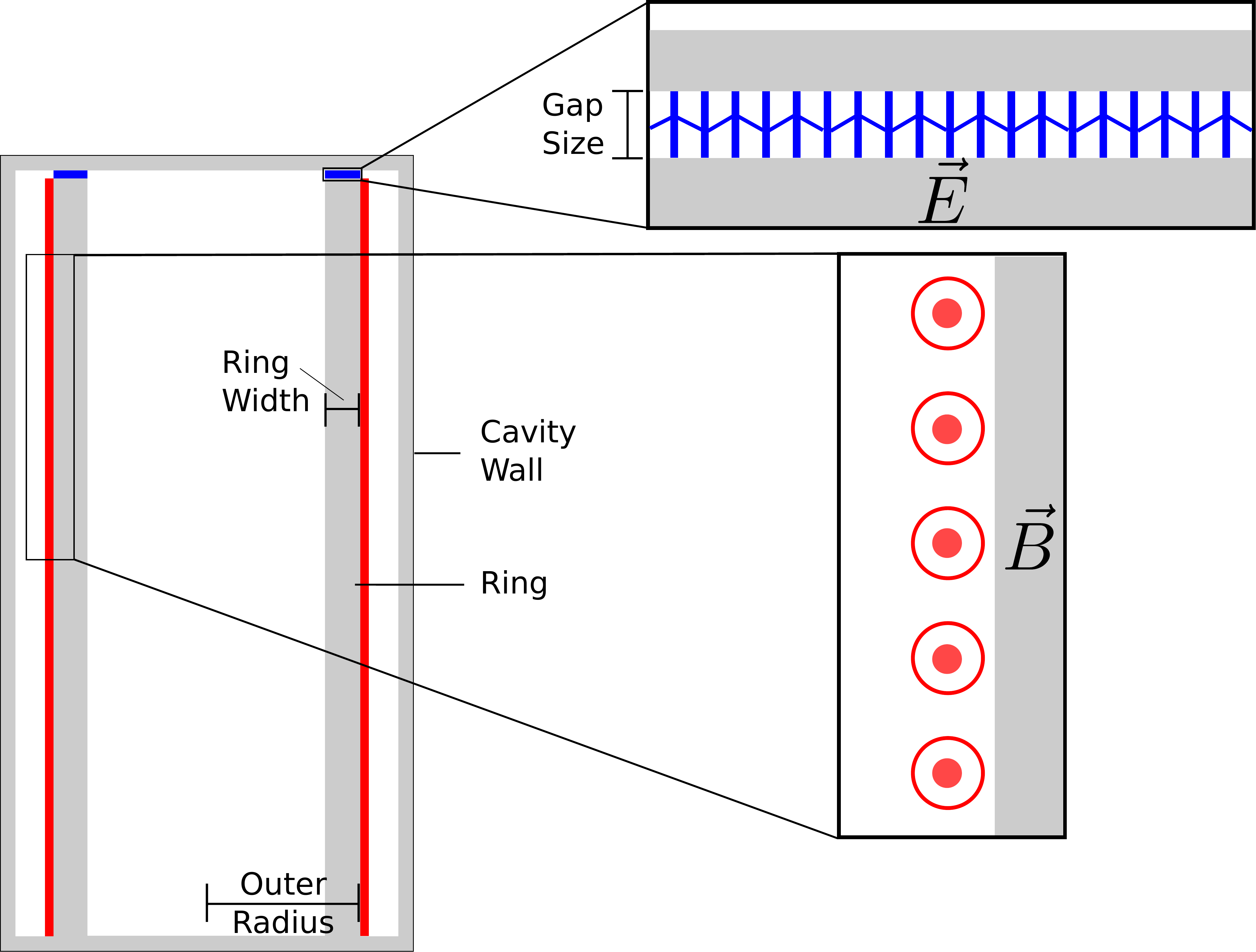}\label{fig:ringcav}}
	\caption{Cross-sections of {\bfseries{(a)}} the ``post" cavity design and {\bfseries{(b)}} the ``ring" cavity design, with electric (blue) and magnetic (red) field locations inset. The electric field propagates axially in the volume between the top of the post or ring and the cavity lid, whereas the magnetic field is in the azimuthal direction near the outside edge of the post or ring.}
\end{figure*}
where $f_a$ is the frequency of the generated photon and $\beta=\frac{v_a}{c} << 1$ for virialized axions \cite{ADMX1}. As $m_a$ is an unknown parameter, it is important to search for axions over a wide range of frequencies. 

To date experiments such as ADMX, a leading haloscope search, have probed regions of the parameter space~\cite{ADMX1,ADMX2,ADMX2011}, and there are several new designs and proposals to explore other ranges~\cite{Baker2012,ADMXHF2014,wispdmx,Seviour2014,orpheus}. Recent work suggested using a magnetometer enhanced by a resonant LC circuit in a magnetic field to try and detect low mass axions ($<$10$^{-7}$~eV)~\cite{LCaxions,SThomas}, other proposals employ LC resonators for the related problem of hidden sector photon dark matter detection~\cite{LCHSP1,LCHSP2}. Low axion mass regimes are difficult to access experimentally with a haloscope as the frequency of haloscope cavity modes is inversely proportional to cavity size, and the available space within strong magnetic fields puts lower limits on the frequencies that can be searched. However, this range is of great theoretical interest as axions originating from a pre-inflationary breaking of the Peccei-Quinn symmetry would manifest in this region~\cite{preinflation}. We propose that 3D resonant LC structures with spatially separated electric and magnetic field components, such as re-entrant cavities~\cite{RRcav,rigorous,tuning,splitring}, are well suited for such low mass Haloscope searches.
\section{Re-entrant cavities}
\begin{figure*}[t!]
	\subfigure[]{
		\includegraphics[width=0.45\textwidth]{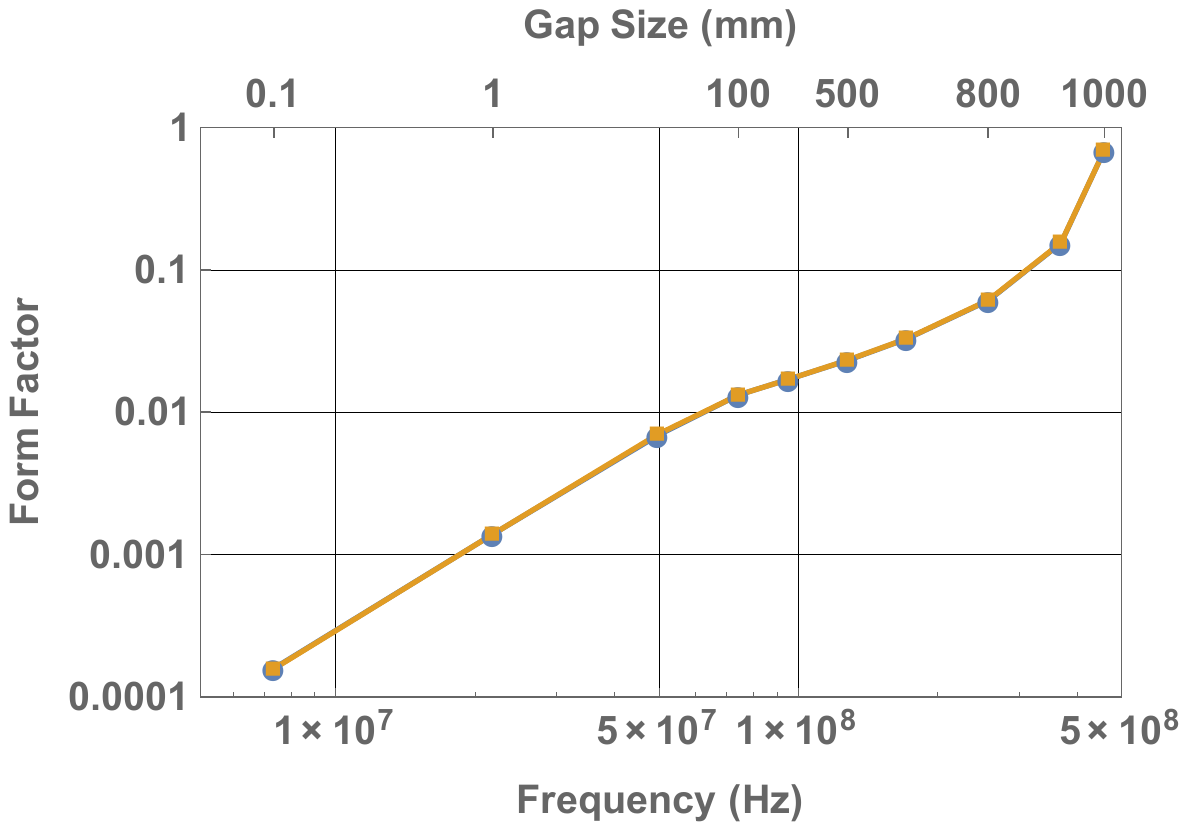}\label{fig:postforms}
	}
	\subfigure[]{
		\includegraphics[width=0.45\textwidth]{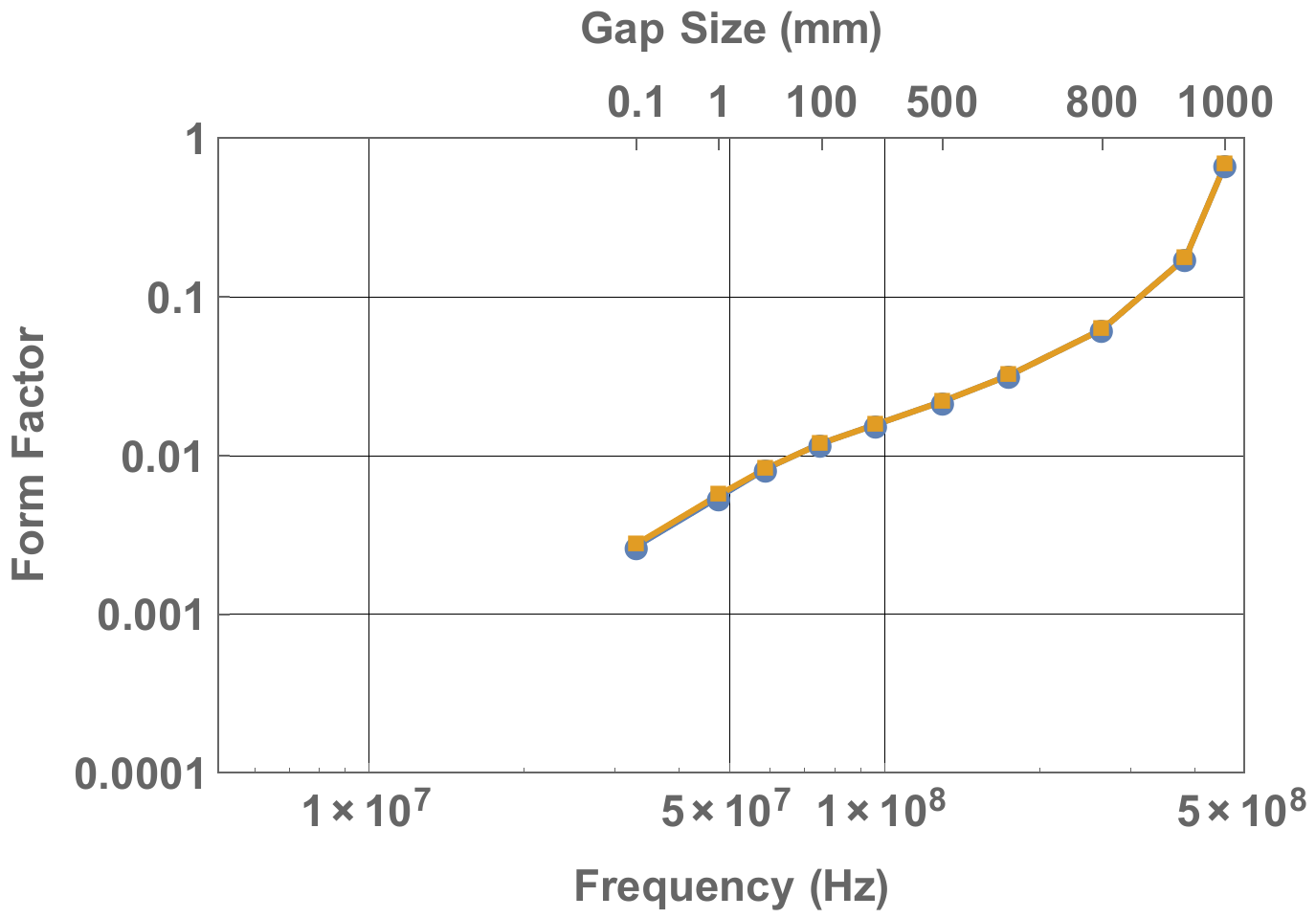}\label{fig:ringforms}
	}
	\caption{Plots of the simulated electric (blue, circles) and magnetic (yellow, squares) form factors vs mode frequency for {\bfseries{(a)}} the ``post" cavity (fig.~\ref{fig:postcav}) and {\bfseries{(b)}} the ``ring" cavity (fig.~\ref{fig:ringcav}). Geometries of these cavities are discussed in section III.}
	\label{fig:forms}
\end{figure*}
Re-entrant cavities are a type of resonant structure which are analogous to an oscillating LC circuit - they usually contain an element such as a metallic or dielectric post or ring, with a small gap between this element and the top or bottom of the cavity. Adjusting the size of this gap adjusts the frequency of the resonant mode. Since in a re-entrant cavity the electric and magnetic fields are spatially separated and highly localized, we must consider the electric and magnetic couplings of the axion independently in order to accurately esitmate the sensitivity. We can consider these couplings by computing the form factor, C~\cite{OurPRL}
\begin{equation}
\text{C} = \frac{\text{C}_\text{E}+\text{C}_\text{B}}{2}
\end{equation}
Where the electric and magnetic form factors $\text{C}_\text{E}$ and $\text{C}_\text{B}$ respectively are given by
\begin{equation}
\text{C}_\text{E}=\frac{\left|\int dV_{c}\vec{E_c}\cdot\vec{\hat z}\right|^2}{V_c\int dV_{c}\mid E_c\mid^2}\label{CE}
\end{equation}
\begin{equation}
\text{C}_\text{B}=\frac{\frac{\omega_a^2}{c^2}\left|\int dV_{c}\frac{r}{2}\vec{B_c}\cdot\vec{\hat\phi}\right|^2}{V_c\int dV_{c}\mid B_c\mid^2}.\label{eq:CB}
\end{equation}
Here $\vec{E_c}$ is the electric field of the cavity mode, $\vec{B_c}$ is the magnetic field of the cavity mode, $\text{V}_c$ is the volume of the cavity and $\omega_a$ is the angular frequency of the photon generated via the Primakoff effect. It is important to note that r and $\phi$ refer to the natural radial and azimuthal coordinates of the cylindrical coordinate system of the solenoid in which the cavity is embedded. Noting the radial dependence in the magnetic form factor of eq.~\eqref{eq:CB}, we will explore the possibility of employing novel re-entrant cavity topologies to ``push" the magnetic field into the outer radial regions of the cavity in an attempt to maximize $\text{C}_{\text{B}}$. Plots of typical electric and magnetic form factors as a function of re-entrant cavity mode frequency are presented in fig.~\ref{fig:forms}.

To model the re-entrant cavities we utilized the COMSOL Multiphysics Software package. Cavities were designed to fit inside the 50~cm diameter ADMX magnet bore~\cite{ADMX1} for the purposes of comparison. The software was used to calculate the cavity volume, geometric factor and electromagnetic form factors, the product of these parameters being the figure of merit for our analysis. The cavity mode dependent geometric factor, G, is given by
\begin{equation}
G=\frac{\omega\mu_0\int\left|\vec{H}\right|^2dV_c}{\int\left|\vec{H}\right|^2dS_c}
\end{equation}
and is related to the cavity mode quality factor by
\begin{equation}
Q=\frac{G}{R_s}.
\end{equation}
Here $\omega$ is the angular frequency of the cavity mode, $\mu_0$ is the permeability of free space, $\vec{H}$ is the magnetic field in the cavity, $\text{V}_\text{c}$ is the volume of the cavity, $\text{S}_\text{c}$ is the surface of the cavity and $\text{R}_\text{s}$ is the surface resistance of the walls of the cavity. The product of form factor, volume and geometric factor was chosen as the figure of merit for our analysis, as upon examination of eq.~\eqref{eq:paxion} it is clear that form factor, volume and quality factor are the only mode dependent factors, the remaining terms in the equation are set by properties of the axion and other components of the experimental setup. For a given static magnetic field, a given axion mass, density and coupling strength the product CVG is directly proportional to the power generated in the haloscope cavity, and thus a good indicator of sensitivity. Formally, the CVG product is defined as\\
\begin{align}
\begin{split}
CVG=&\left(\frac{\left|\int dV_{c}\vec{E_c}\cdot\vec{\hat z}\right|^2}{2\int dV_{c}\mid E_c\mid^2}+\frac{\frac{\omega_a^2}{c^2}\left|\int dV_{c}\frac{r}{2}\vec{B_c}\cdot\vec{\hat\phi}\right|^2}{2\int dV_{c}\mid B_c\mid^2}\right)\\
&\times\frac{\omega\mu_0\int\left|\vec{H}\right|^2dV_c}{\int\left|\vec{H}\right|^2dS_c}
\end{split}
\end{align}
These re-entrant designs were engineered to push the resonant frequency lower than what could be achieved in an equivalently sized TM or TE mode, enabling the possibility of searching for lower mass axions with pre-existing cavity Haloscope infrastructure.
\section{Design and simulation results}
\begin{figure}[b!]
	\includegraphics[width=0.9\columnwidth]{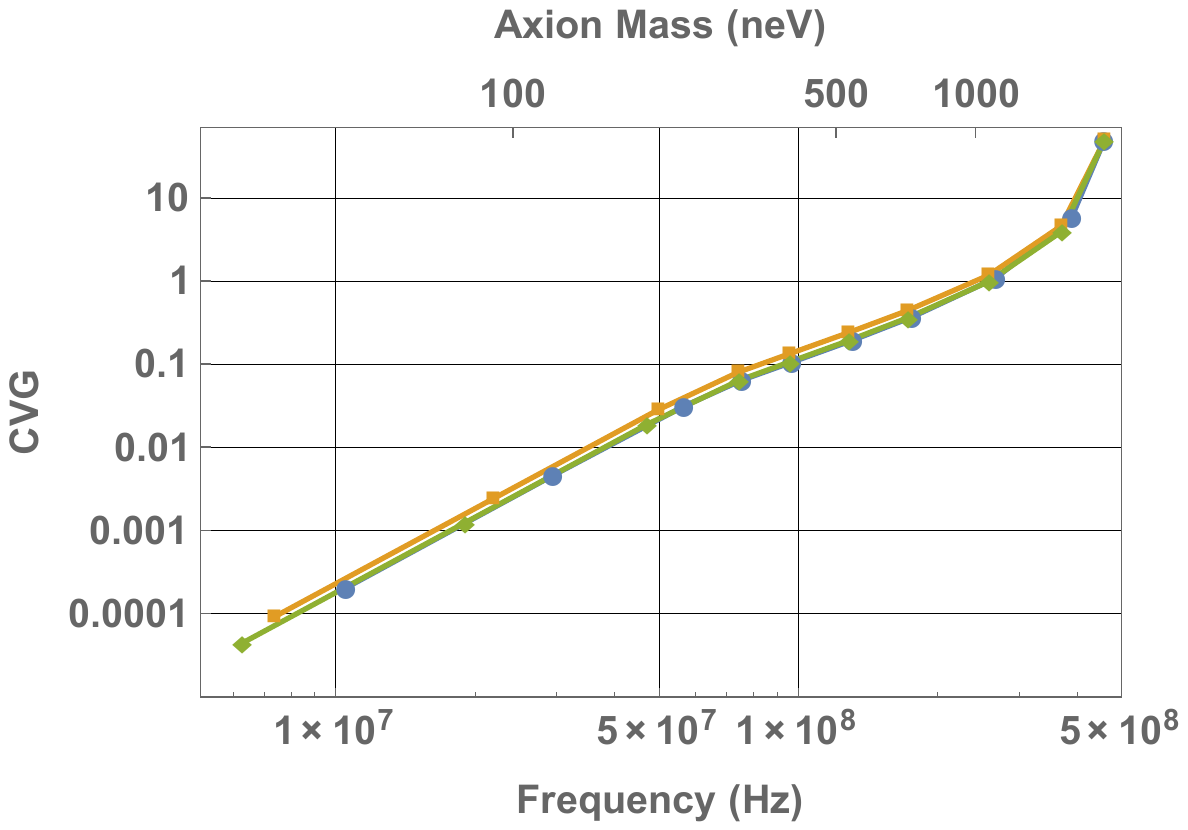}
	\caption{A plot of the figure of merit, CVG (product of form factor, volume and geometric factor, which is proportional to Q-factor) versus cavity mode frequency for the ``post" cavity design (fig.~\ref{fig:postcav}) with post radii of 5 cm (blue, circles), 9 cm (yellow, squares) and 15 cm (green, diamonds), corresponding to 20, 36, and 60\% of the cavity radius respectively.}
	\label{fig:postradii}
\end{figure}
\begin{figure*}[t!]
	\subfigure[]{
		\includegraphics[width=0.346436\textwidth]{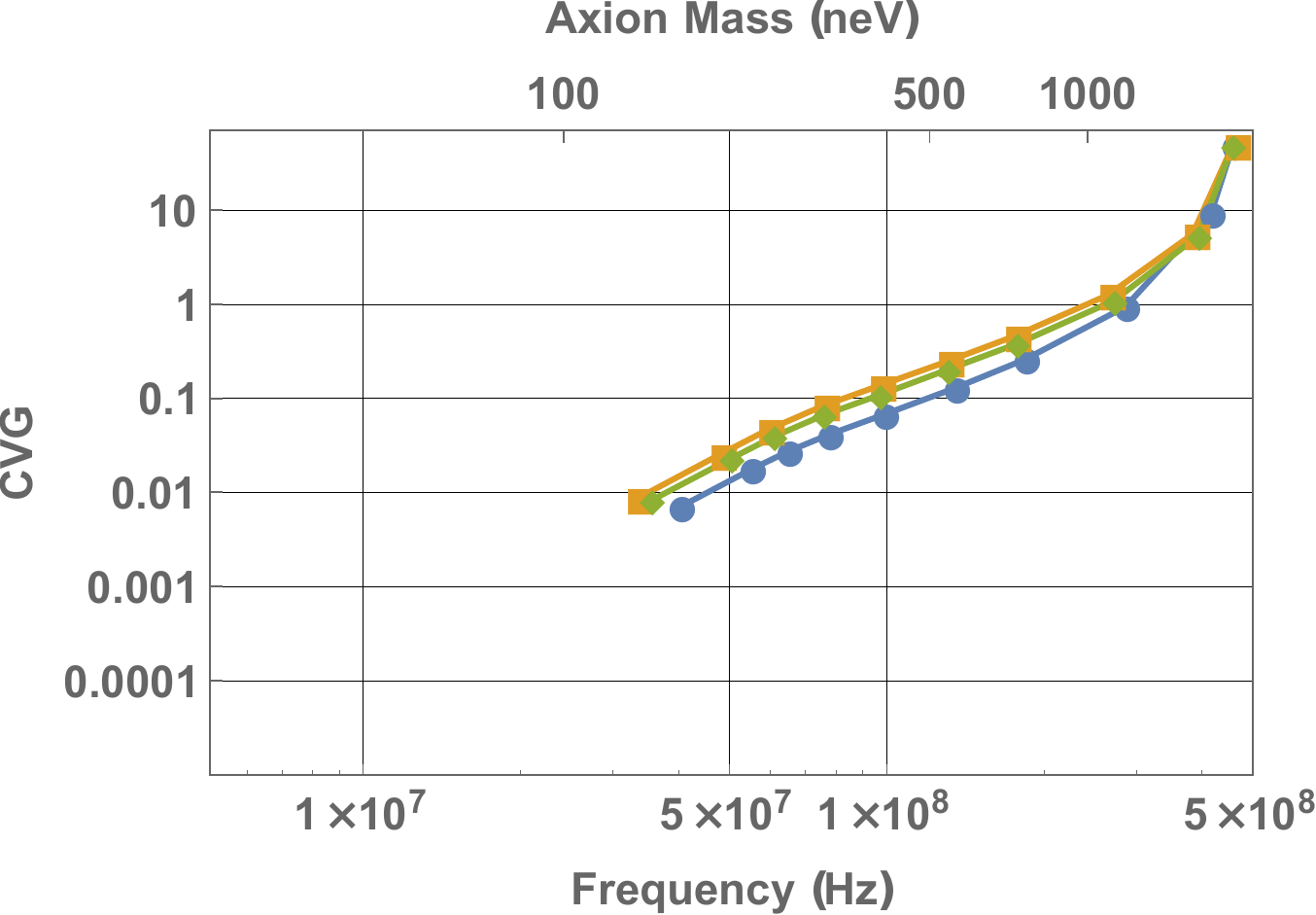}\label{fig:constthick}
	}
	\subfigure[]{
		\includegraphics[width=0.603564\textwidth]{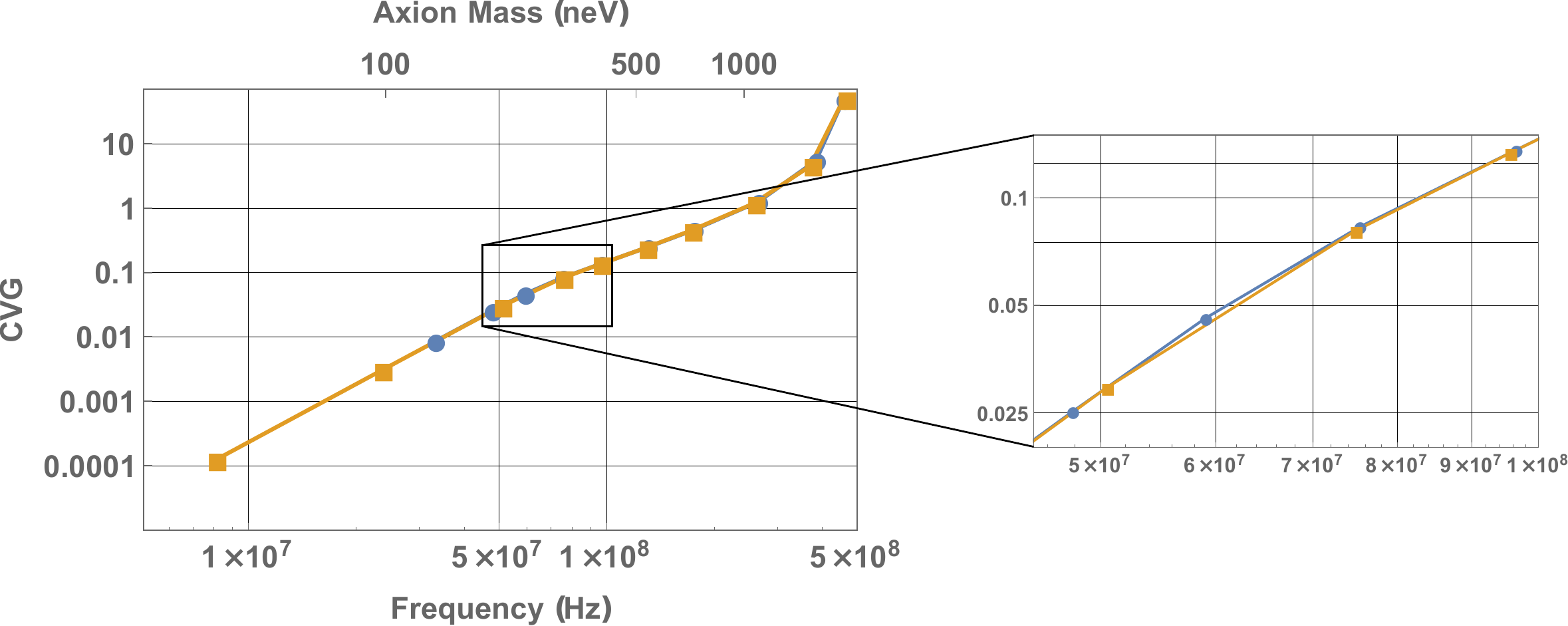}\label{fig:constrad}
	}
	\caption{Plots of the figure of merit, CVG, versus cavity mode frequency for the ``ring" cavity design (fig.~\ref{fig:ringcav}) with {\bfseries{(a)}} a constant ring thickness of 1 mm, with outer radii of 2.5 cm (blue, circles) 9 cm (yellow, sqaures) and 15 cm (green, diamonds) (corresponding to 10, 36, and 60\% of the cavity radius respectively) and {\bfseries{(b)}} a constant outer radius of 10 cm (40\% of the cavity radius), with ring thicknesses of 1 mm (blue, circles) and 5 cm (yellow, squares). Note that plot {\bfseries{(b)}} contains an inset showing a smaller selection of the total scanning range in order to illustrate the slight improvement in sensitivity offered by the lower ring thickness.}
\end{figure*}
A design utilizing a solid central post as a tuning mechanism was considered, see fig.~\ref{fig:postcav}. Such a design is readily tunable by adjusting the gap between the end of the post and the lid of the cavity. Cavities with posts between 5 mm and 22.5 cm in radius (2 and 90 \% of total cavity radius respectively) were modelled inside the ADMX 50 cm diameter, 1 m length magnet bore in order to determine the effect of post radius on sensitivity to axions and cavity mode tuning range. It was generally found that larger post radii were favourable, as seen in fig.~\ref{fig:postradii}, which shows a selection of the simulated data. Higher post radii offer an improved tuning range and initially offer better sensitivity when compared with lower post radii, however diminishing returns in sensitivity occur for post radii beyond $\sim$9 cm, or $\sim$36\% of the cavity radius. Consider a cavity of diameter 50 cm and length 1 m, containing a central post of radius 9 cm.

Such a cavity embedded in the ADMX magnet bore is sensitive to axions and tunable in the region $\sim$30-2000 neV (corresponding to frequencies of $\sim$7.4-480 MHz) when the size of the gap between the top of the post and the lid of cavity increases from 100~$\mu$m to 1 m. The electric and magnetic form factors range from $\sim$0.00016 at the lower mass to $\sim$0.69 at the higher mass (see fig.~\ref{fig:postforms}).

In order to utilize more of the cavity volume, and increase sensitivity at the higher end of this range, a cavity design utilizing a thin central ring was considered (see fig.~\ref{fig:ringcav}). Such a cavity is also readily tunable by adjusting the gap between the top of the ring and the lid of the cavity. Cavities were modelled with varying ring thicknesses, between 1 mm and 5 cm, and varying outer radii, between 1 and 20 cm (4 and 80 \% of cavity radius respectively). It was found that for a given outer radius, higher ring thicknesses offered larger tuning ranges, but lower sensitivity than lower ring thicknesses.  Furthermore, for a given ring thickness, sensitivity increased up to a certain point, much like the ``post" cavity scheme (see figs.~\ref{fig:constthick} and \ref{fig:constrad} which show a selection of the simulation data). When exploring varying outer radii diminishing returns in sensitivity occurred, again, for values greater than $\sim$9 cm, or 36\% of the cavity radius.

\begin{figure*}[t]
	\subfigure[]{
		\includegraphics[width=0.42\textwidth]{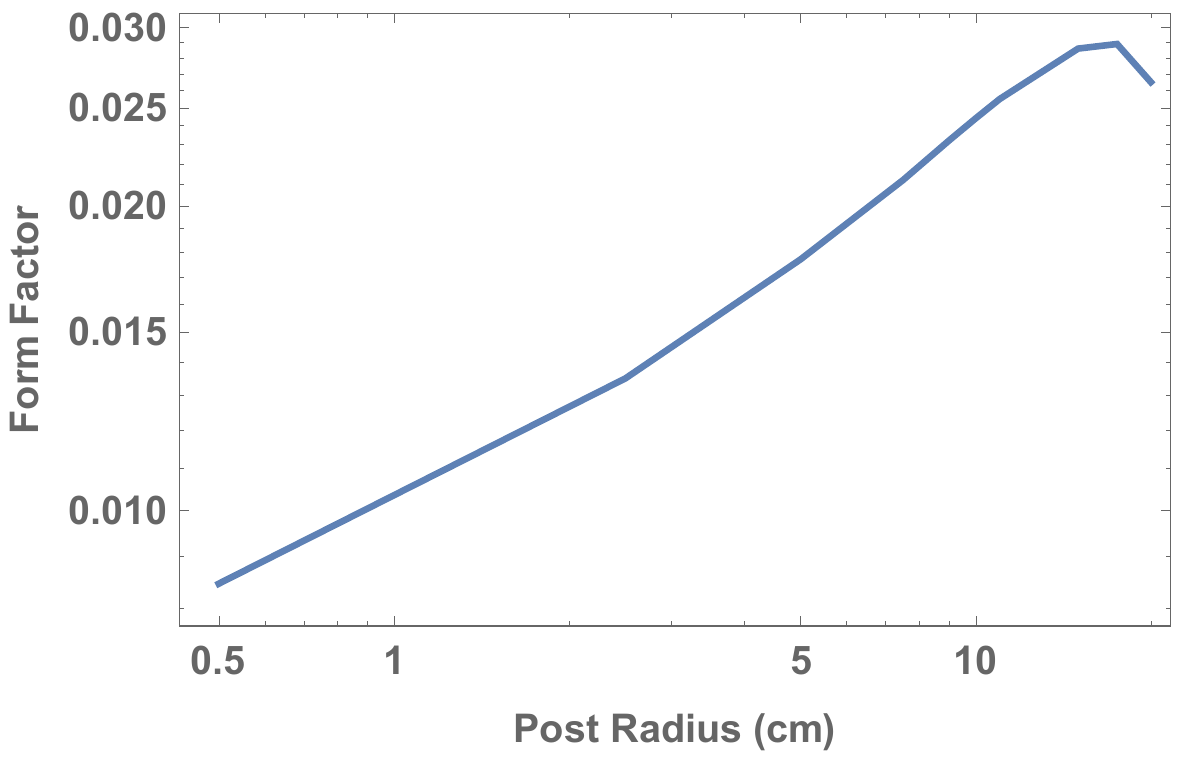}
	}
	\subfigure[]{
		\includegraphics[width=0.42\textwidth]{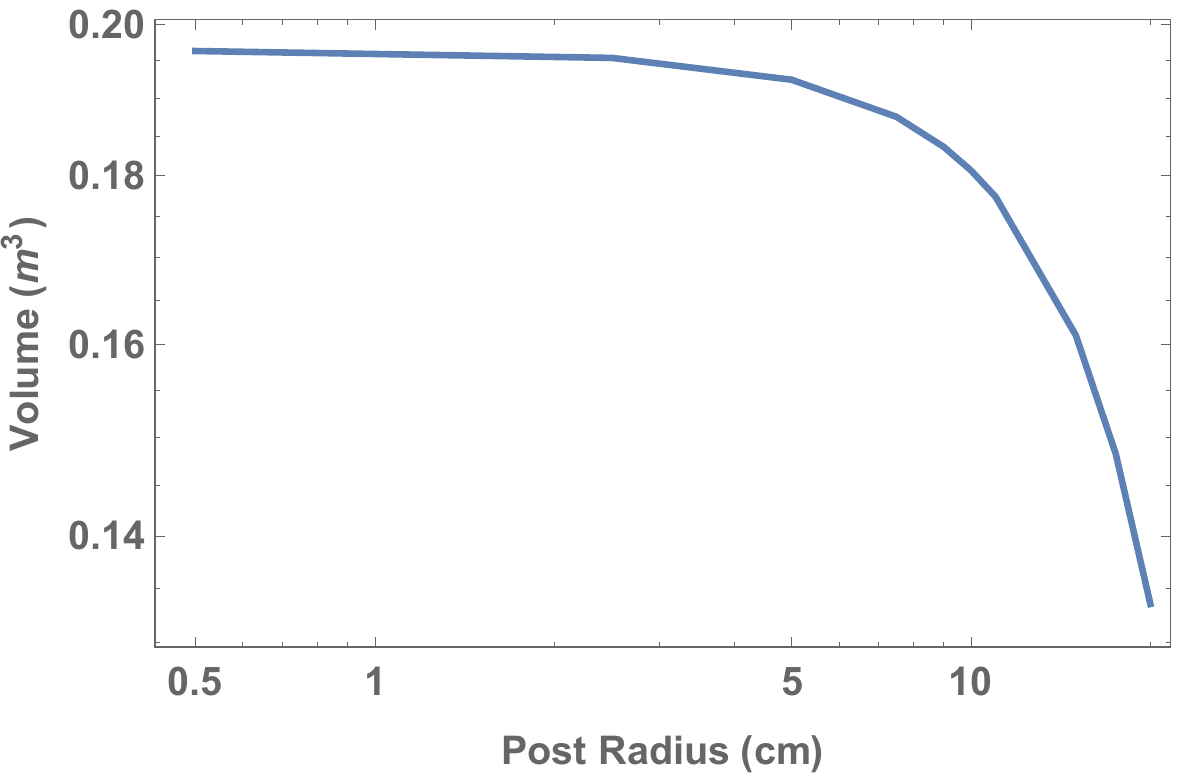}
	}	\subfigure[]{
		\includegraphics[width=0.42\textwidth]{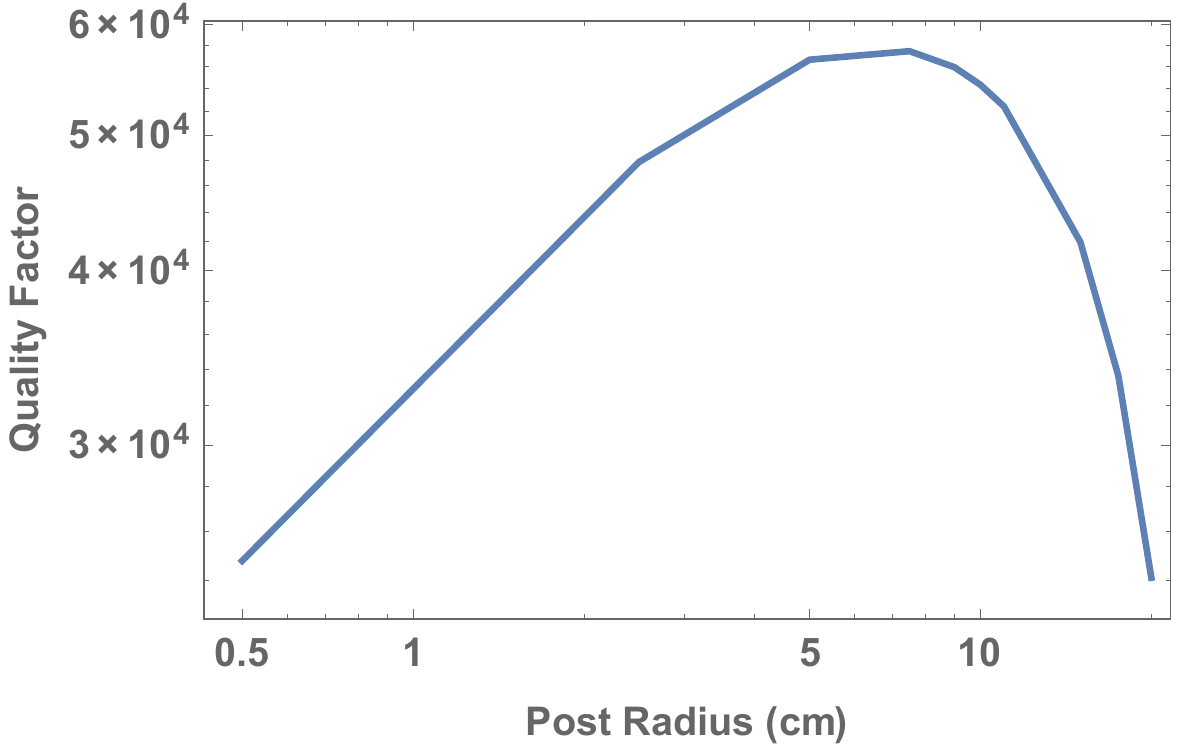}
	}
	\subfigure[]{
		\includegraphics[width=0.42\textwidth]{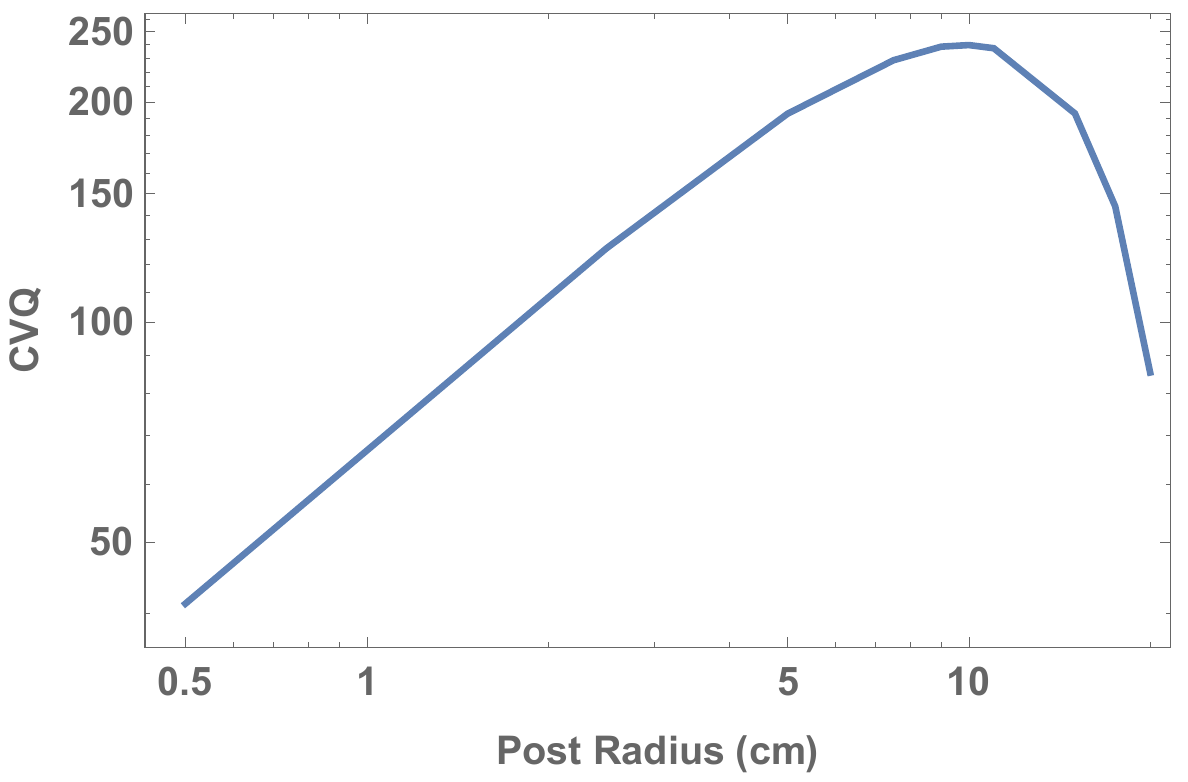}
	}
	\caption{Plots of {\bfseries{(a)}} full electromagnetic form factor, {\bfseries{(b)}} cavity volume, {\bfseries{(c)}} mode quality factor, and {\bfseries{(d)}} the product thereof as a function of post radius for a constant gap size of 50 cm.}
	\label{fig:illustrate}
\end{figure*}

Consider a cavity of diameter 50 cm and length 1 m, containing a central ring of outer radius 9 cm and thickness 1 mm. Such a cavity embedded in the ADMX magnet bore is sensitive to axions and tunable in the region $\sim$136-2000 neV (corresponding to frequencies of $\sim$33-480 MHz) when, again, the size of the gap between the top of the ring and the lid of cavity increases from 100~$\mu$m to 1 m. The electric and magnetic form factors range from $\sim$0.0028 at the lower mass to $\sim$0.69 at the higher mass (see fig.~\ref{fig:ringforms}). It is worth noting that as the gap size between the top of the post or ring and the cavity lid increases, the mode structure gradually transitions from a re-entrant mode to a typical TM mode, and indeed the final mode structure is the $\text{TM}_{010}$ mode employed by the ADMX collaboration. See \cite{RCtoTM} for a rigorous discussion of this process.\\
The reduction in sensitivity for post or ring radii beyond 36~\% of the cavity radius in each case may seem counterintuitive but is in fact unsurprising. As post radii increase, the magnetic and electric form factors both increase (due to increased radial dependence and also a larger electric mode volume), however, total cavity volume and geometric factor decrease. Initially, the gains in form factor outweigh the losses in volume and geometric factor, but this reverses for radii beyond $\sim$36\% of cavity radius. This is illustrated in fig.~\ref{fig:illustrate}.\\
It is further worth noting that the stated results for maximum sensitivity are something of a simplification. Different post and ring radii and thicknesses offer different tuning ranges, and in different parts of the parameter space different cavity geometries offer optimal sensitivity. The 36\% radius value is quoted as this provides optimal sensitivity over a large section of the tuning range, and in an experiment it is more practical to have one or a small number of cavity geometries. However, even when a different geometry is optimal the value of radius does not deviate much from 36\%, and maximum sensitivity is almost always achieved with radii in the range 25\% to 36\% of the cavity radius. With radii beyond $\sim$36\% of the cavity radius sensitivity declines rapidly.
\section{Sensitivity and outlook}
\begin{figure}[t]
	\includegraphics[width=0.99\columnwidth]{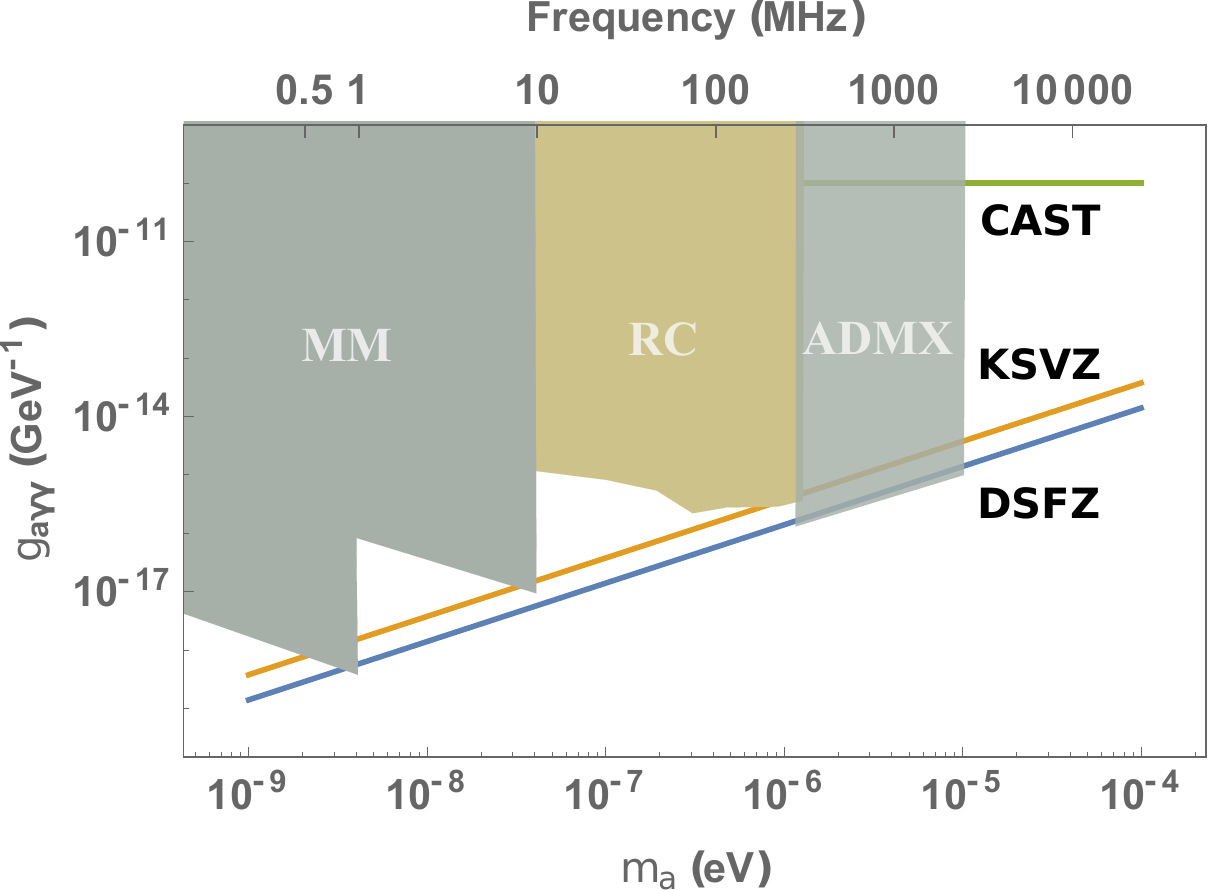}
	\caption{Predicted axion-photon coupling exclusion limits for the re-entrant cavity designs, using the assumptions outlined in the main text (RC). Current and future bounds from ADMX and CAST are shown for comparison, along with the predicted exclusion limits for magnetometer experiments (MM) presented in \cite{LCaxions}.}
	\label{fig:reexc}
\end{figure}
We will now consider the sensitivity of an axion haloscope experiment utilizing these 3D lumped LC resonators in the ADMX magnet bore and spanning the region $\sim$35-1300 neV, from the lowest frequencies achievable with such resonators, up to the existing lower bounds of the ADMX experiment. Optimal sensitivity in the range $\sim$35-136 neV is achieved with the ``post" type cavity with a post radius of 6 cm (although larger posts up to 9 cm provide better sensitivity over the upper end of the tuning range, in this low frequency range the 6 cm post provides the best sensitivity), where as optimal sensitivity in the range $\sim$136-1300 neV is achieved with the ``ring" type cavity with a ring outer radius of 9cm and a ring thickness of 1 mm. Assuming that the relevant resonator for the frequency is embedded in the ADMX 8~T magnet bore, a surface resistivity of copper of $\sim$1~m$\Omega$ at 20 mK and 100 MHz \cite{cryocopper}, a local galactic dark matter density of $\sim$0.5 GeV/$\text{cm}^3$ (typically assumed in previous searches \cite{ADMX1,ADMX2}), and a quantum limited SQUID-based amplifier with an effective noise temperature of $\sim$15~mK, one could constrain the axion-photon coupling constant with a Signal-to-Noise Ratio (SNR) of 2 to the level shown in fig.~\ref{fig:reexc} with a total measurement time of $\sim$935 days. This is a very promising search region, and lumped 3D LC resonator experiments are capable of spanning almost all of the viable parameter space between the proposed magnetometer experiments~\cite{LCaxions} and the various searches of the ADMX collaboration~\cite{ADMX1,ADMX2,ADMX2011} with adequate sensitivity. Higher sensitivity could be achieved by narrowing the search region, or increasing the total scan time.

It is important to note that this proposed search is in a lower frequency range than current haloscopes, thus the noise backgrounds may exhibit some difference. There are many more ambient and laboratory sources of RF noise in the MHz regime compared to the GHz regime, but provided that sufficient radiation shielding is employed in the cryogenic system the impact of these sources should be minimal. Additionally, at lower frequencies the quantum noise limit of amplifiers decreases, which could lead to gains in sensitivity. Furthermore, the lower end of the MHz regime could be directly sampled by low-noise high frequency analogue-to-digital converters, which would eliminate the need to mix down the output signal, thus removing a potential source of noise.

In conclusion lumped 3D LC resonators, commonly known as re-entrant cavities, can be employed as sensitive axion haloscopes in low mass ranges currently unprobed in the parameter space. By employing novel re-entrant cavity topologies one can maximize the sensitivity of such experiments, utilizing both the electric and magnetic coupling of axions to photons. Furthermore, these techniques push the cavity frequency into ranges inherently lower than are achievable in the equivalent size of empty cavity, and can be readily implemented in existing haloscope infrastructure.
\begin{acknowledgements}
This work was supported by Australian Research Council grants CE110001013, as well as the Australian Postgraduate Award and the Bruce and Betty Green Foundation.\\
\end{acknowledgements}

\end{document}